\pdfoutput=1

\documentclass[letterpaper, 10 pt, conference]{ieeeconf}  

\IEEEoverridecommandlockouts                              

\usepackage{url}
\usepackage{hyperref}
\usepackage{microtype}
\usepackage{graphicx}
\usepackage{booktabs} 
\usepackage{float}
\usepackage[utf8]{inputenc}
\usepackage{graphics,graphicx}
\usepackage{amsfonts}
\usepackage{amsmath}
\usepackage{amsthm}
\usepackage{amssymb}
\usepackage{comment}
\usepackage{mathtools}
\allowdisplaybreaks
\theoremstyle{definition}
\newtheorem{lemma}{\textbf{Lemma}}

\newtheorem{theorem}{\textbf{Theorem}}

\usepackage{hyperref}
\usepackage[caption = false]{subfig}
\usepackage{bbding}
\usepackage{adjustbox}
\usepackage{stix}
\usepackage{babel}
\usepackage{caption}
\renewcommand{\implies}{\;\Rightarrow\;}

\usepackage{color}

\title{
\vspace{-2cm}
\scriptsize 2021 IEEE. Personal use of this material is permitted. Permission from IEEE must be obtained for all other uses, in any current or future media, including reprinting/republishing this material for advertising or promotional purposes, creating new collective works, for resale or redistribution to servers or lists, or reuse of any copyrighted component of this work in other works.\\ \vspace{1.5cm}
\LARGE \bf 
Towards cyber-physical systems robust to communication delays: A differential game approach
}

\author{Shankar A. Deka$^{1}$, Donggun Lee$^{2}$ and Claire J. Tomlin$^{1}$
\thanks{This work has been submitted to the IEEE for possible publication. Copyright may be transferred without notice, after which this version may no longer be accessible.}
\thanks{$^{1}$Shankar A. Deka and Claire J. Tomlin are with the Department of Electrical Engineering and Computer Sciences, University of California, Berkeley, 2594 Hearst Ave, Berkeley, CA 94720, USA.
        {\tt\small deka.shankar@berkeley.edu, tomlin@eecs.berkeley.edu}}%
\thanks{$^{2}$Donggun Lee is with the Department of Mechanical Engineering, University of California, Berkeley, 6141 Etcheverry Hall, Berkeley, CA 94720, USA.
        {\tt\small donggun\_lee@berkeley.edu}}%
\thanks{This work is supported by NIFA, by the DARPA Assured Autonomy program, and by the ONR BRC program in multibody systems.}
}

\begin{document}

\maketitle
\thispagestyle{empty}
\pagestyle{empty}

\begin{abstract}
Collaboration between interconnected cyber-physical systems is becoming increasingly pervasive. Time-delays in communication channels between such systems are known to induce catastrophic failure modes, like high frequency oscillations in robotic manipulators in bilateral teleoperation or string instability in platoons of autonomous vehicles. This paper considers nonlinear time-delay systems representing coupled robotic agents, and proposes controllers that are robust to time-varying communication delays. We introduce approximations that allow the delays to be considered as implicit control inputs themselves, and formulate the problem as a zero-sum differential game between the stabilizing controllers and the delays acting adversarially. The ensuing optimal control law is finally compared to known results from Lyapunov-Krasovskii based approaches via numerical experiments. 
\end{abstract}

\section{Introduction}

\begin{figure*}[!hbt]
    \centering
    \quad
    \subfloat[Bilateral teleoperation.]{\includegraphics[width=0.35\textwidth]{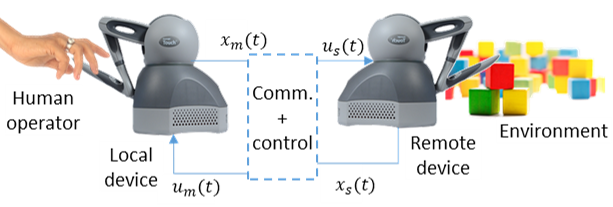}}
    \;
    \subfloat[String stability in vehicle platoons (Figure adopted from \cite{Feng2019String}).]{\includegraphics[width=0.3\textwidth]{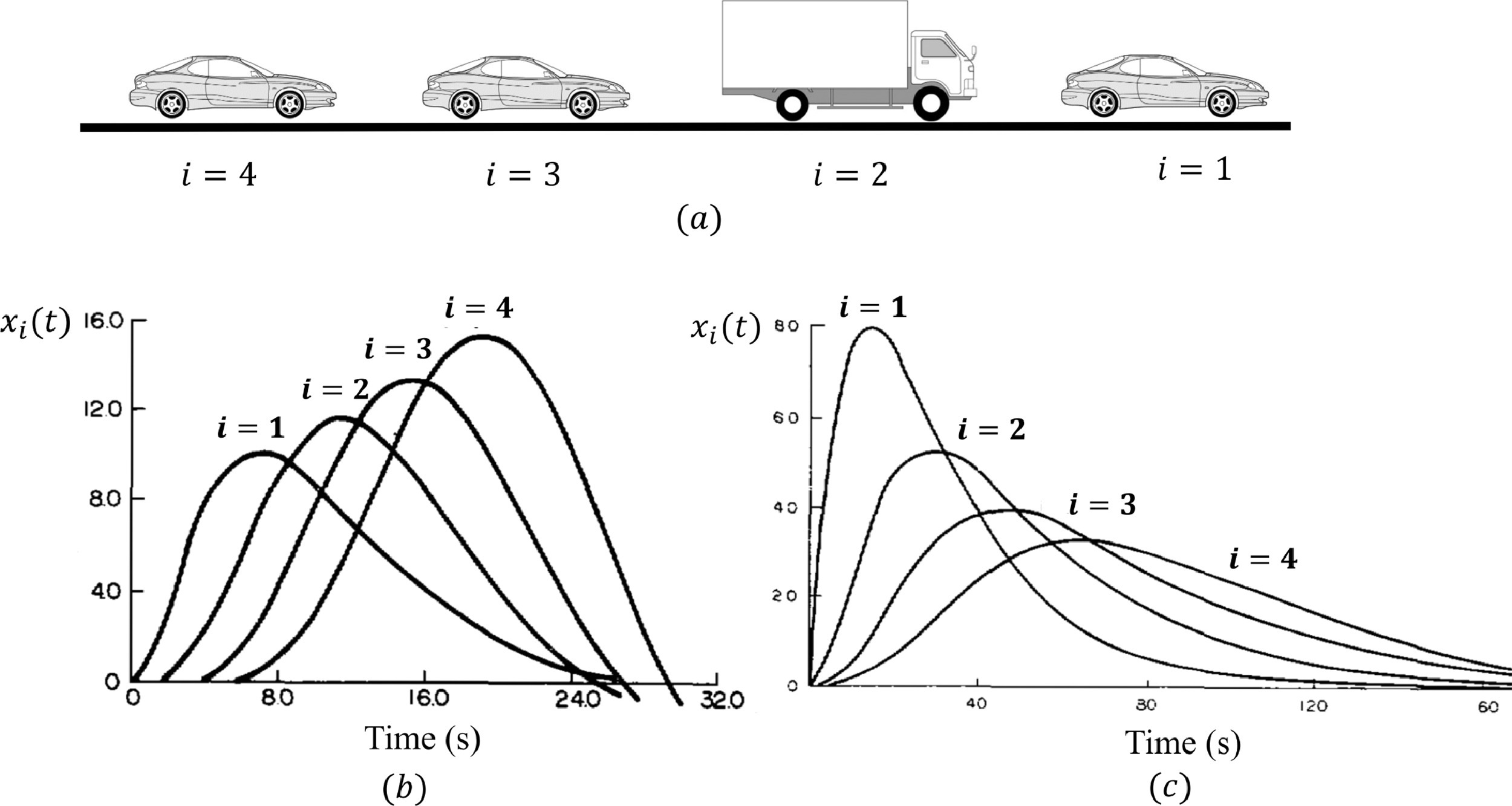}}\;
    \subfloat[Formation control.]{\includegraphics[width=0.25\textwidth]{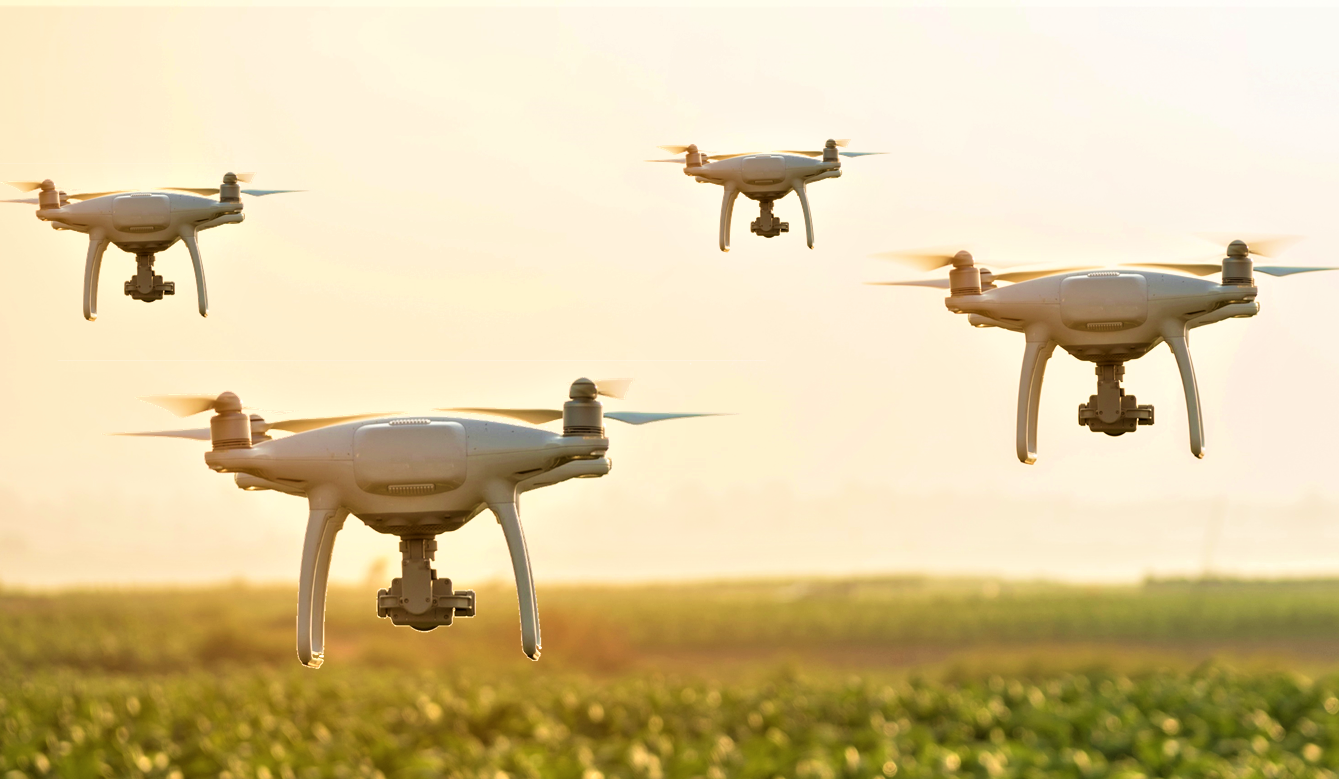}}
    \caption{Some delay sensitive CPS applications.}
    \label{fig:my_label}
\end{figure*}

Time-delays are an inevitable part of systems that involve flow of information, and may arise, for example, due to processing sensor measurements before using them in decision and control, communication between multiple robotic agents in a distributed system, or simply due to inherent transient lags in a system's response. Although the time delay model we choose in this paper is general enough to cover any of these scenarios, we are mainly motivated by cyber-physical systems (CPS) applications involving two or more interconnected agents exchanging state information among one-another to achieve a control task such as multi-UAV formation control, autonomous vehicular platooning \cite{liu2001effects}\cite{ besselink2017string}\cite{ qin2019experimental}, bilateral teleoperation \cite{nuno2009position}\cite{deka2019stable} or load-frequency control in power grids \cite{lou2019learning}\cite{xiahou2020robust}\cite{ganesh2021learning}. For a general review of time-delays systems, control methods and applications, please see \cite{liu2019survey}\cite{richard2003time}.

One of our objectives in this paper is to study how time-varying delays can adversely destabilize a system that is originally stable in the absence of any delays. We intend to construct optimal time-delays that maximize the instability over a finite time horizon, as well as stabilizing controllers which take the time-delay information as part of their feedback strategy. Though this may seem simple, the implicit manner in which time-delays enter a dynamical system and affect its behavior, makes it a non-trivial problem. In a sense, the delays control the rate of flow of information, like measurements received by a feedback-controller, rather than controlling the state trajectories directly. In fact, optimal control/controllability with respect to time delays is still an open problem: The current literature has addressed only a limited part of the problem, where the delays are constant valued and/or \textit{a priori} fixed, and are not the main object of control. For example, we refer to \cite{salamon1984controllability}\cite{Basin2007Optimal}\cite{basin2008optimal} for linear systems, and more recent work on nonlinear systems \cite{kirchner2019level}\cite{plaksin2020minimax}\cite{plaksin2021viscosity}. In this paper, although we do not attempt to tackle this problem in its full generality, we introduce an approximation to nonlinear time-delay systems (TDS) that allows us to conveniently handle time-varying delay signals. With our approximate dynamics, we formulate the problem of designing robust feedback controllers for TDS as a multi-player differential game, and to that end, we explore a few recently developed computational tools rooted in optimal control theory.

The contributions of this paper are organized into the following sections. We begin by presenting the general problem and control objective as described by a functional differential equation, and motivate our paper through a practical example in Section \ref{sec:main}. The next section sets up the delay-robustification goal in a game-theoretic framework, which is made tractable through simplifying approximations introduced later in Section \ref{sec:approx}. Following this, we solve our control problem, as well as identify safe sets via backward-reachability computations described in Section \ref{sec:HJ}. In Section \ref{sec:approx_bounds}, we present essential analysis to answer how close our approximate solution is to the original problem. Additional numerical experiments and concluding remarks are presented in the final two sections.

\section{Main problem}\label{sec:main}
We start by describing our problem in continuous time through delay-differential equations. Let us consider a nonlinear system of the form
\begin{align}
\label{eq:TDS}
    \begin{split}
        \dot{x}_1 &= f_1(x_1(t),x_2(t-d_2(t)),u_1(t)),\\ 
        \dot{x}_2 &= f_2(x_2(t),x_1(t-d_1(t)),u_2(t)),
    \end{split}
\end{align}
with asymmetric, time-varying delays $d_1 , d_2 \in [0,d_{\max}]$ for $d_{\max}\in\mathbb{R}_+$, states $x_1,x_2\in\mathbb{R}^n$, and control inputs $u_1,u_2 \in\mathbb{R}^m$.
This may physically represent a system of two robotic agents, exchanging their state information for, say synchronization. Since we are primarily interested in robustness with respect to time-delays $d_1,d_2$, we shall assume that without the delays and control, the nominal system\footnote{For the rest of the paper, we shall use the term ``nominal" to refer to systems with $u_i(t),d_i(t) \equiv 0$.} is designed to be stable, and may satisfy other properties such as $\mathcal{L}_2,\mathcal{L}_\infty$ boundedness of velocities, asymptotic convergence of the states $x_1$ and $x_2$, forward-invariance with respect to some `safe-set' etc., that are typically expected in a robotic system. Even for such otherwise well-behaved nominal systems, injection of small delays are known to adversely degrade their behavior. We would like to robustify these nominal systems against delay-induced instabilities using feedback controls $u_1 = u_1(x_1, x_2(t-d_2(t)), t, d_2(t))$ and $u_2 = u_2(x_2, x_1(t-d_1(t)), t, d_1(t))$, for any bounded delays $d_1(t)$ and $d_2(t)$.

To illustrate the TDS \eqref{eq:TDS} and control challenges introduced by time-delays, let us consider bilateral teleoperation of robotic manipulators as a specific example.

\noindent\textbf{Example 1.} The system dynamics for robotic manipulators is commonly modelled using Euler-Lagrange equations:
\begin{align}\label{eq:teleoperation}
\begin{split}
&M_1(\theta_1)\Ddot{\theta}_1 + C_1(\theta_1,\dot{\theta}_1)\dot{\theta}_1 + G_1(\theta_1) = \tau_1 \\ 
&M_2(\theta_2)\Ddot{\theta}_2 + C_2(\theta_2,\dot{\theta}_2)\dot{\theta}_2 + G_2(\theta_2) = \tau_2
\end{split}
\end{align}
where $\tau_1,\tau_2$ are the joint torques and vectors $\theta_1,\theta_2$ are generalized joint coordinates. In order to achieve position coordination between the two manipulators, a common choice is to pick the inputs as
\begin{align}\label{eq:control_delay}
\begin{split}
\tau_1 &= -k_1(\theta_2(t-d_2(t)) - \theta_1(t)) + b_1\dot{\theta}_1\\
\tau_2 &= -k_2(\theta_1(t-d_1(t)) - \theta_2(t)) + b_2\dot{\theta}_2.
\end{split}
\end{align}
Thus taking $x_i = [\theta_i,\dot{\theta}_i]$, the closed-loop system can be written compactly as equation \eqref{eq:TDS} (with $u_1,u_2=0$). In the absence of time-delays, this controller can be easily shown to asymptotically stabilize the system (assuming gravity compensation), and achieves position coordination for all values of parameters $k_1,k_2,b_1,b_2>0$. Delays of even a small magnitude can disrupt this behaviour, and the controller gains need to be chosen more carefully, as described by the following known result.
\begin{theorem}\label{th:bilat}\cite{nuno2009position} (Asymptotic stabilization):
\textit{
 For closed loop system given by equations \eqref{eq:teleoperation} and \eqref{eq:control_delay}, $\|\theta_1 - \theta_2\| \in \mathcal{L}_{\infty}$ and $\dot{\theta}_1,\dot{\theta}_2 \in \mathcal{L}_2\cap\mathcal{L}_{\infty}$ for any bounded time-delays $d_1(t)$ and $d_2(t)$ if the inequality 
\begin{equation}\label{eq:sufficient}
4b_1b_2 \ge (T_1^{*2} + T_2^{*2})k_1k_2
\end{equation}
is satisfied, where $T^*_1$ and $T^*_2$ are the upper bounds on delays $d_1(t)$ and $d_2(t)$, respectively. Moreover, as $t \rightarrow \infty$, $\theta_1 \rightarrow \theta_2$ and $\dot{\theta}_1,\dot{\theta}_2\rightarrow 0.$ }
\end{theorem} 

\begin{figure}[!b]
    \centering
    \vspace{-0.5cm}
    \includegraphics[width=0.8\columnwidth]{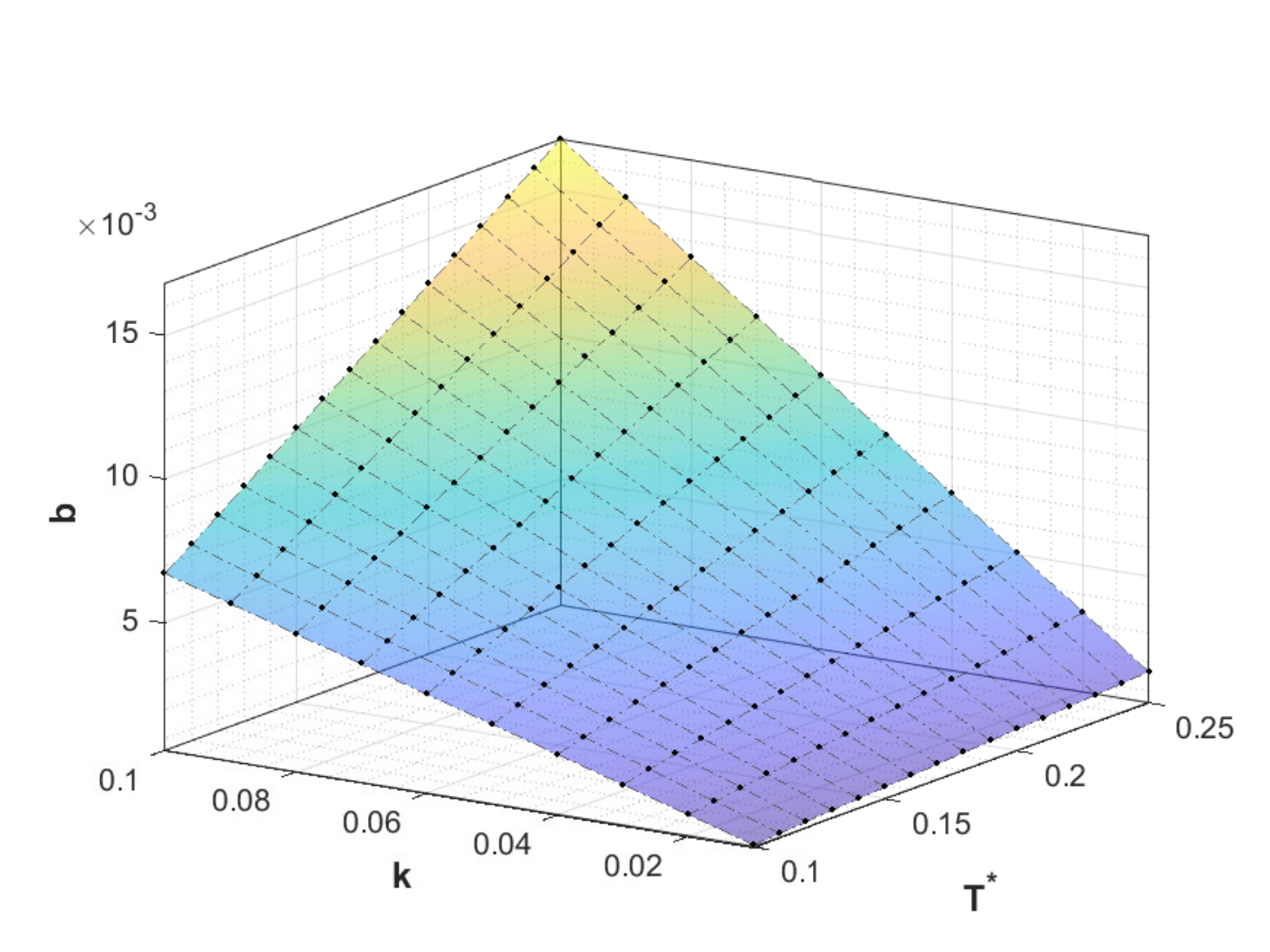}
    \caption{For the system \eqref{eq:teleoperation}, with $M_{1,2} = I, C_{1,2}=0, G_{1,2}=0$, the above surface shows values of parameters that do not satisfy equation \eqref{eq:sufficient}, and yet represent stable systems, with $(\theta_1 - \theta_2)\rightarrow 0$ for appropriate sinusoidal delays. One can see that with larger values of $k$, the system is more sensitive to time-delays, as the slope with respect to $T^*$ is much steeper at $k=0.1$ than it is at $k=0.02$. Since this sufficient condition for stability is agnostic to exact delay signals and considers all possible delays lying within a fixed upper bound, the result tends to be conservative. In particular, the high damping terms bring robustness at the expense of performance, as observed through experiments in our prior work \cite{deka2019stable}. }
    \label{fig:stable}
\end{figure}

The sufficient condition in Theorem \ref{th:bilat} is obtained by appropriately constructing Lyapunov-Krasovskii functionals, and the controller gains tend to be conservative (please see Figure \ref{fig:stable} and its description), as is common with a Lyapunov-based design paradigm. Furthermore, such a result holds asymptotically, which means that on a finite time horizon, the performance can be poor for appropriately selected time-delays acting as adversarial disturbances. Here, poor performance could mean that dissipation of oscillations and convergence to zero position errors are slow).

We shall illustrate this point in detail later once we have a way to obtain adversarial time-delays that optimally disrupt the system performance by maximizing some cost function over a finite time horizon. But another motivating argument for considering optimal control design is that we can make use of state and delay information as feedback for tighter stabilizing controllers, rather than designing policies that are delay independent, such as in \cite{nuno2009position}\cite{deka2019stable} and prior works.
\section{Differential game formulation and reachability analysis for time-delay systems}\label{sec:HJ}
In order to formalize our broad control objective of robust control against time-varying delays, we shall formulate our problem as a multi-player game between the time-varying delays and the stabilizing control inputs.
For some given system  $\dot{x}=f(x,u,d),$
let us define a cost function
\small
\begin{equation}\label{eq:cost_}
J(t,x(\cdot),u(\cdot),d(\cdot)) = \int_t^T l_t(t,x(s),u(s),d(s))ds + l_T(x(T)),
\end{equation}
\normalsize
where $u(\cdot) \in\mathcal{U}(t)\coloneqq\{u:[t,T]\rightarrow \mathbb{U}~|~u \text{ is measurable}\}$, and  $d(\cdot)\in\mathcal{D}(t,d_{\max})\coloneqq\{d :[t,T]\rightarrow [0,d_{\max}]~|~d \text{ is measurable} \}$.
The objective is to solve 
\begin{equation}
\label{eq:costJ}
\max_{\delta\in\Delta(t)} \min_{u\in \mathcal{U}(t)}   J(t,x(\cdot),u(\cdot),\delta[u(\cdot)]),
\end{equation}
where $\Delta(t)$ is the set of non-anticipative strategies \cite{elliott1972existence}, which is a reaction function for the second player with respect to the first player's action without using future information:
\begin{align*}
    \Delta(t,d_{\max})\coloneqq&\{\delta:\mathcal{U}(t)\rightarrow\mathcal{D}(t,d_{\max}) ~| ~ \forall s\in[t,\tau] \text{ and } \notag\\ &u,\bar{u}\in\mathcal{U}(t),\text{      if } u(\tau)=\bar{u}(\tau) \text{ a.e. } \tau\in[t,s],\notag\\
    &\text{then } \delta[u](\tau)=\delta[\bar{u}](\tau) \text{ a.e. } \tau\in[t,s] \}.
\end{align*}

The corresponding Hamilton-Jacobi-Isaacs (HJI) equation is 
\begin{equation}\label{eq:HJI_}
0 = \frac{\partial V}{\partial t} + \min_u \max_d \left\{ l_t(t,x,u,d) + \frac{\partial V}{\partial x}\cdot f(x,u,d) \right\},
\end{equation}
with initial condition $V(x,T) = l_T(x)$. Reachability analysis using the HJI-PDE involves the use of a very specific cost function; one where stage cost is zero and terminal cost is the signed distance function from some target final set $\left\{x \;|\; h(x)=0 \right\}$. The optimal value function $V$ evaluated at $t=0$ describes the \textit{backward-reachable set}\footnote{Backward-reachable set with respect to a target set defines as the set of initial states from where the target set can be reached over some fixed time horizon by applying an appropriate control. Please see \cite{bansal2017hamilton} for details.} in terms of the zero-level set of $V$. Typically, these are obtained by numerically solving the HJI-PDE, although closed-form solutions may be found in special cases.

\section{Control-affine approximation for delays}\label{sec:approx}
Before we can solve the problem \eqref{eq:cost_}-\eqref{eq:HJI_}, we need an alternative representation of the TDS \eqref{eq:TDS}, since the delays enter the dynamics in a manner that is not amenable for analysis or control synthesis. Since, $\hat{x}_i(t) \coloneqq  x_i(t - d_i(t))\approx  x_i(t) - \dot{x}_i(t - d_i(t))\cdot d_i(t)$, one may approximate the equation \eqref{eq:TDS} as,
\begin{eqnarray*}\label{eq:affine}
\dot{x}_1 &=& f_1(x_1,\hat{x}_2,u_1)\\ 
\dot{x}_2 &=& f_2(x_2,\hat{x}_1,u_2)\\ 
\dot{\hat{x}}_1 &=& \dot{x}_1(t-d_1(t))\big(1-\dot{d}_1\big) \approx \big(x_1 - \hat{x}_1\big)w_1\\
\dot{\hat{x}}_2 &=& \dot{x}_2(t-d_2(t))\big(1-\dot{d}_2\big) \approx \big(x_2 - \hat{x}_2\big)w_2\\ 
\dot{d}_2 &=& 1 - d_2w_2\\
\dot{d}_1 &=& 1-d_1w_1
\end{eqnarray*}
With these new auxiliary states and inputs, we define $\tilde{X}=[x_1,x_2,\hat{x}_1,\hat{x}_2, d_1,d_2]$ and $\tilde{U} = [u_1,u_2,w_1,w_2].$ This can be made more precise by considering more terms in the Taylor expansion of $\hat{x}_i(t)$ as in \cite{zhang2020input}. If the dynamics is linear and delays are constant, other transformations can be applied to further eliminate the delays \cite{fiagbedzi1986feedback}. A concise review can be found in \cite{richard2003time}, particularly relevant is section 6.3.

Such a simplification comes at the cost of increased state dimension which can be detrimental to computational algorithms that scale exponentially with the state dimension\footnote{This is also commonly referred to as the `curse of dimensionality'.}. We thus present another way to approximate the TDS \eqref{eq:TDS}, without augmenting the state-space, as follows.
\begin{align} \label{eq:simple}
\begin{split}
    &\dot{x}_1 = y_1\\ 
    &\dot{x}_2 = y_2\\
    \text{Subject to:}& \\
\end{split}\\
    \label{eq:constraints}
\begin{split}
    &y_1 = f_1(x_1,x_2 - y_2d_2,u_1) + w_1\\ 
    &y_2 = f_2(x_2,x_1 - y_1d_1,u_2) + w_2,\\
\end{split}
\end{align}
where $w_1$ and $w_2$ represent the error terms, belonging to the set $\mathcal{W} \coloneqq \{ w:[0,T]\rightarrow \mathbb{R}^n ~|~w\text{ is measurable} \}$.
The algebraic constraints \eqref{eq:constraints} in general are implicit, which means $y_{1},y_{2}$ need to be solved numerically. However, for control-affine robotic systems, they can be obtained in closed-form (as $y_1 = y_1(x_1,x_2,d_2,u_1,w_1), \, y_2 = y_2(x_2,x_1,d_1,u_2,w_2)$) in a straightforward manner.

It is important that we take a close look at the tightness of such an approximation, since the stabilizing controllers designed using system \eqref{eq:simple}-\eqref{eq:constraints} will have to perform just as well with the original TDS \eqref{eq:TDS}. We shall discuss the approximation bounds on the value function later, but first we consider the error bounds on the dynamics.
\begin{lemma}\label{lemma:approx_error1}
\textit{
    Given the two systems \eqref{eq:TDS}, and \eqref{eq:simple}-\eqref{eq:constraints}, let us assume that partial derivatives of functions $f_1,f_2$ with respect to the second argument are bounded. Then, the approximation errors $w_1,w_2$ are bounded. Furthermore, $\|w_1\|,\|w_2\|\rightarrow 0$ as $\|d_1\|,\|d_2\| \rightarrow 0$. In particular,
    \begin{align}
    \label{eq:lip_bound}
    \begin{split}
    \|w_1\| &\le L_1|o(\|d_2\|)|\\
    \|w_2\| &\le L_2|o(\|d_1\|)|
    \end{split}
    \end{align}
for some constants $L_1,L_2>0$.}
\end{lemma}
\noindent \textbf{Proof.} Please see Appendix \ref{subsec:Proof_approx_error1}.\\

Now that we have a more convenient way to handle delays, we can study how time-delays affect the system stability and performance, and overall safety. In order to do so, we employ the HJI approach described in the last section for control design and analysis. But before that, we illustrate our approximation on a simple time-delay model:\\

\noindent\textbf{Example 2.} (Time-delayed double integrator ) For the states $x_1=[v_1,p_1]$ and $x_2=[v_2,p_2]$, consider the dynamics
\begin{align}\label{eq:another}
\begin{split}
    \dot{v}_1 &= \tau_1 = -k(p_1 - p_2(t-d_2)) + bv_1 + u_1\\
    \dot{p}_1 &= v_1\\
    \dot{v}_2 &= \tau_2 = -k(p_2 - p_1(t-d_1)) + bv_2 + u_2\\
    \dot{p}_2 &= v_2.
    \end{split}
\end{align}
The terms $p_i(t-d_i)$ can be approximated as $p_1(t-d_1) \approx p_1(t) - v_1(t)d_1 + \frac{1}{2}\tau_1d_1^2$ and $p_2(t-d_2) \approx p_2(t) - v_2(t)d_2 + \frac{1}{2}\tau_2d_2^2$. Substituting these in equation \eqref{eq:another}, we obtain $\tau_1$ and $\tau_2$ in terms of the delays $d_1$ and $d_2$ by the system of linear equations:
\begin{eqnarray}\label{eq:polynomial_delay}
 &\tau_1 - \frac{kd_2^2}{2}\tau_2 \approx -k(p_1 - p_2 + v_2d_2) + bv_1 + u_1 \\ \nonumber
 &\tau_2 - \frac{kd_1^2}{2}\tau_1 \approx -k(p_2 - p_1 + v_1d_1) + bv_2 + u_2,
\end{eqnarray}
Solving the two simultaneous equations would give us $\tau_1,\tau_2$ approximately as polynomial functions in $d_1$ and $d_2$, and linear in states $x_1$ and $x_2$. The closed-loop dynamics for system \eqref{eq:another} can then be written as 
\begin{eqnarray}\label{eq:Approx}
 \dot{x} &=&  f_p(x,u,d) = A_p(d)x + B_p(d)u,
\end{eqnarray}
where $x=[x_1^T,x_2^T]^T$ is the augmented state vector, $u=[u_1,u_2]^T$ is the control input, and $A_p(d)$ and $B_p(d)$ are matrices with entries that are polynomial in $d=[d_1,d_2]$. \qed\\

One can solve then solve the HJI-PDE \eqref{eq:HJI_} numerically for this system, as we do for our results in Section \ref{sec:Exp}. The resulting control policy is demonstrated to effectively stabilize under time-delays later in Figure \ref{fig:Stabilization}.

\section{Tightness of our approximation \eqref{eq:simple}-\eqref{eq:constraints}}\label{sec:approx_bounds}


Our ultimate goal is to design optimal controllers $u_1$ and $u_2$ that allows the TDS \eqref{eq:TDS} to perform robustly in presence of time-varying delays $d_1$ and $d_2$. In order to make our problem tractable, we rewrite it in the form of equations \eqref{eq:simple} and \eqref{eq:constraints}. The terms $w_1$ and $w_2$ however, are treated as uncertain quantities that are dependant on the delays, and are bounded using the Lipschitz constants $L_1$ and $L_2$. As $d_1,d_2 \rightarrow 0$, the terms $w_1,w_2 \rightarrow 0$, but for larger values of $d_1,d_2$ we shall consider them as bounded uncertain inputs, and find optimal stabilizing controllers $u_1^*,u_2^*$ in presence of bounded disturbances $d_1,d_2,w_1,w_2$.
We use $\Gamma$ to denote a set of non-anticipative strategies for $w$ with respect to the control $u$ and time delay $d$, which satisfies inequality \eqref{eq:lip_bound} in Lemma \ref{lemma:approx_error1}.

We need to bound the discrepancy between our approximation and the actual TDS dynamics, since the optimal robust controllers computed using the approximate system dynamics are expected to stabilize the original system. Towards that goal, we present the following.

\begin{theorem} \label{th:approx_error2}
    Let us consider a cost functional that is only dependent on the state trajectories:
\begin{equation}
    J(t,x(\cdot)) = l_T(x(T)) + \int_t^T l_s(s,x(s))ds.
\end{equation}
For a control signal and strategies $u$, $\gamma$, $\delta$,
let us further denote the trajectories of systems \eqref{eq:TDS} and \eqref{eq:simple} starting from some point $x$ at some time $t$ as $\phi_a(x,\tau,u(\cdot),\delta[u](\cdot))$ and $\phi_b(x,\tau,u(\cdot),\delta[u](\cdot),\gamma[u,\delta[u]](\cdot))$,
respectively, for $t \le \tau \le T$. Then,
\begin{small}\begin{equation}\label{eq:J_comparision}
     \inf_{\gamma\in\Gamma}\sup_{\delta\in\Delta(d_{\max})}   \inf_u J(t,\phi_b) \le  \sup_{\gamma\in\Gamma} \inf_u J(t,\phi_a) \le  \sup_{\gamma\in\Gamma}\sup_{\delta\in\Delta(d_{\max})}  \inf_u J(t,\phi_b).
\end{equation}\end{small}

Further assume that the dynamics and the stage cost are Lipschitz continuous in the state and control, and the terminal cost is Lipschitz continuous in the state. Then,
\small
\begin{gather}\label{eq:cost_comp}
    \lVert \inf_{\gamma\in\Gamma} \sup_{\delta\in\Delta (d_{\max})} \inf_u J(t,\phi_b) 
    - \sup_{\gamma\in\Gamma} \sup_{\delta\in\Delta(d_{\max}) }
    \inf_u J(t,\phi_b) \rVert \leq C d_{\max}
\end{gather}
\normalsize
for some constant $C$.
\end{theorem}
\noindent\textbf{Proof.} Please see Appendix \ref{subsec:proof_approx_error2}.

From the second inequality in equation \eqref{eq:J_comparision}, it clear that a stabilizing controller $u^*$ for the approximate dynamics \eqref{eq:simple}-\eqref{eq:constraints} will also stabilize the original TDS \eqref{eq:TDS} (assuming minimization of an appropriately chosen cost $J$ over a finite time-horizon $T$ leads to stabilization). This upper bound is the optimal cost under the worst disturbance and modeling error in dynamics \eqref{eq:simple}-\eqref{eq:constraints}, which means that the backward-reachable set computation using our approximate system will yield a conservative estimate of the original system.
From a safety perspective, a conservative design, though not ideally desired, is safe nonetheless. Equation \eqref{eq:cost_comp} establishes that our approximation error between the two costs remains small for small delays.



\section{Numerical experiments and discussion}\label{sec:Exp}
A key point that remains to be demonstrated is that the control and analysis performed using the approximate system \eqref{eq:simple}-\eqref{eq:constraints} can indeed be successfully applied to the actual TDS \eqref{eq:TDS} where the delays influence the dynamics implicitly. In all of the experiments in this section, the optimal stabilizing control $u^*$ is obtained using the approximate system but is tested on the original TDS dynamics. In order to realistically simulate the effects of time delays for our experiments, we use the \textit{variable transportation delay block} in Simulink \cite{documentationsimulation}. 

\begin{figure}[!htp]
    \includegraphics[width=0.9\columnwidth, clip = true, trim = 0cm 6cm 2cm 0cm]{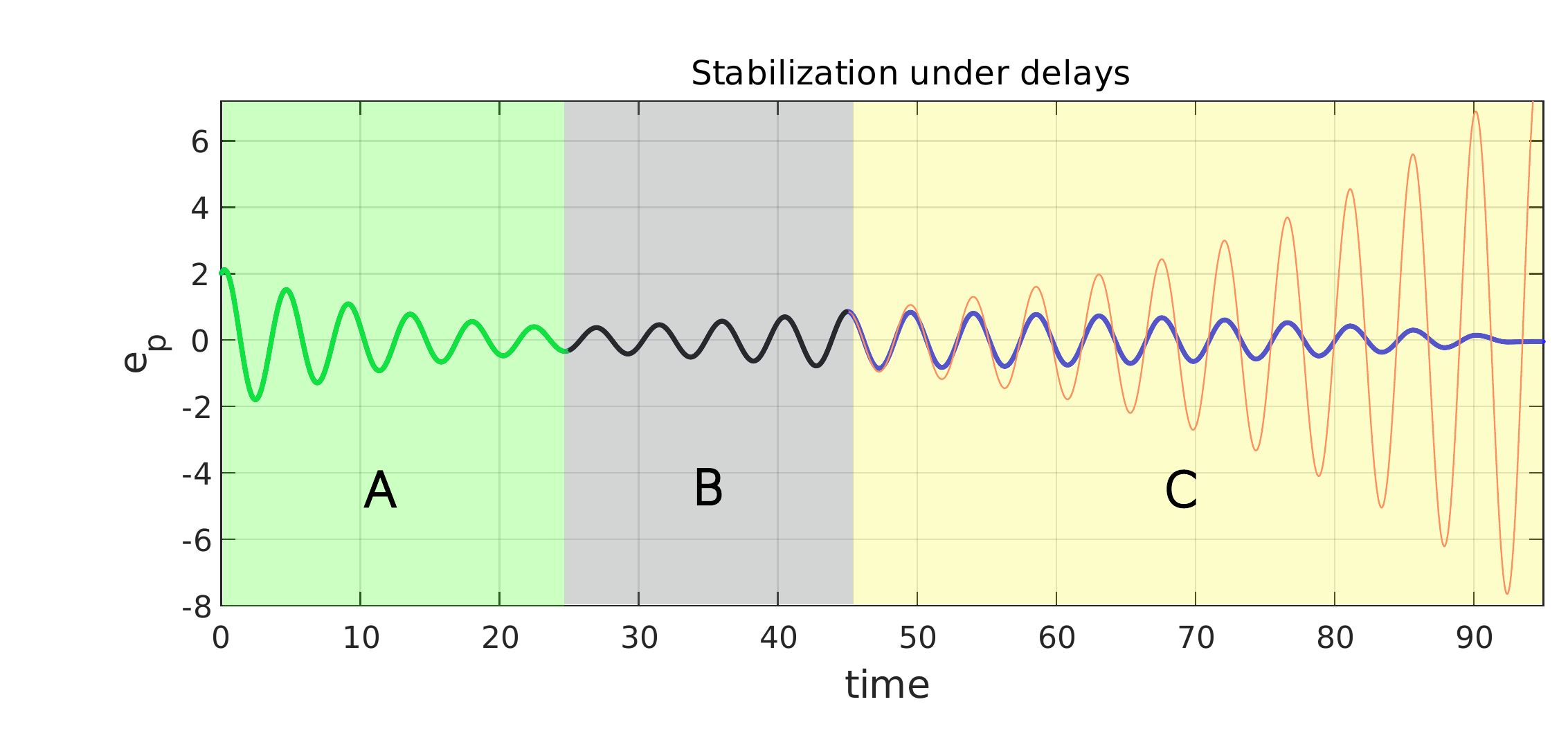}
    \caption{We start again with two double integrator systems with maximum delay $T^*$, and parameters $k,b = 1,0.15$. With $T^*$ set to zero for the first 25 seconds (i.e., no delays), the system oscillations are stable and the position error $e_p \coloneqq p_1 - p_2$ is seen converging to zero in green region `A'. At t=25 seconds, a delay with $T^* = 0.25$ is introduced. The portion of the $e_p$ trajectory in region `B' can be clearly seen to start destabilizing, with gradual increase in oscillation amplitude. At time t=45, we introduce our stabilizing control, which dampens the oscillations in $e_p$ (blue curve) throughout region `C'  until it converges to zero after t=90. Shown in red is evolution of $e_p$ when we do not apply our stabilizing control and allow the oscillations in region `B' to continue growing.}
    \label{fig:Stabilization}
\end{figure}

We use the system described in Example 2 for illustration. Since this model is a special case of Example 1, the sufficient Lyapunov-Krasovskii stability condition of Theorem \ref{th:bilat} still holds. We choose our model parameters to ensure that the delays in fact destabilize the system, and demonstrate the effectiveness of our computed $u^*$ in Figure \ref{fig:Stabilization}.

\begin{figure}[!htp]
    \centering
    \subfloat[]{
    \includegraphics[width=0.6\columnwidth]{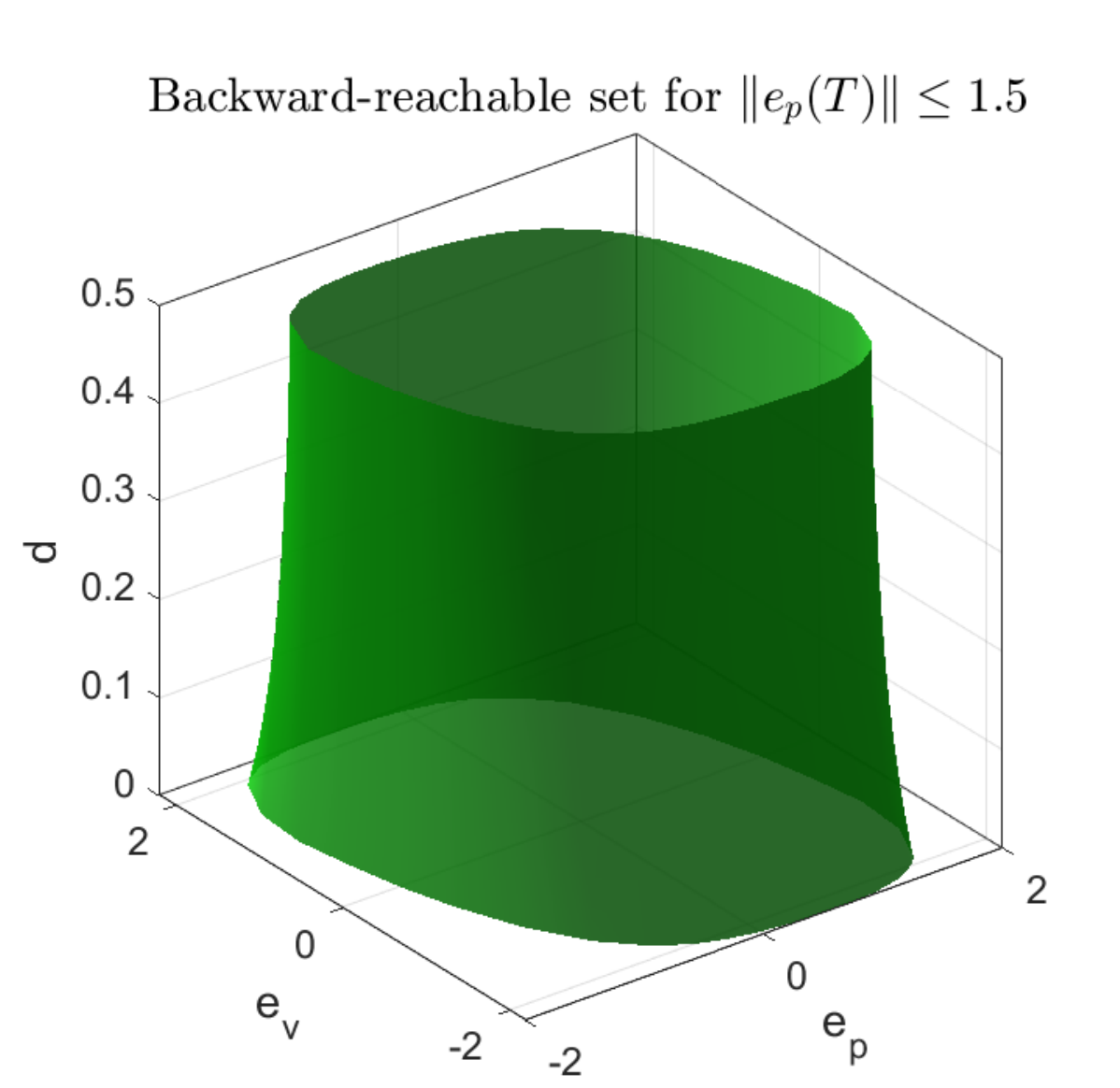}}\\
    \vspace{-0.2cm}
    \subfloat[]{
    \includegraphics[width=0.7\columnwidth]{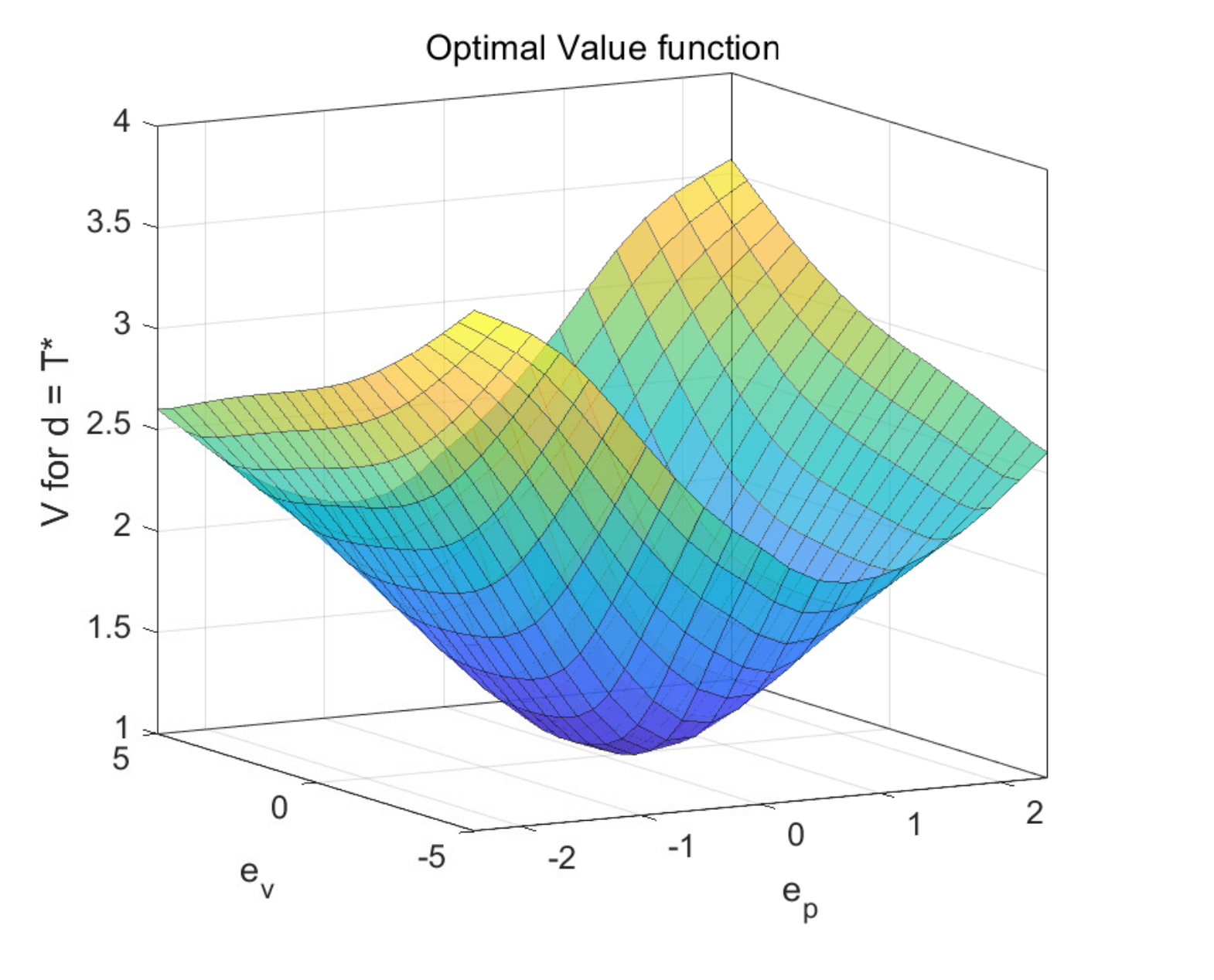}}
    \caption{Reachability of the `optimal' closed-loop system (with optimal controller $u^*$) driven by $d$. (a) The green enclosed region describes the set of initial conditions $(e_p(0),e_v(0),d(0))$ from which the optimal $u^*$ can robustly lead the $e_p(t)$ trajectories, to a bounded region ($\|e_p\|\le1.5$) in finite-time $T$ (chosen to be 10 seconds) for any delay $d(t)$ bounded by maximum delay $T^*=0.5$. One can notice that for larger initial delays, the corresponding set of safe $(e_p,e_v)$ gets smaller. (b) The optimal value function $V(e_p,e_v, d=0.5)$ for points $(e_p,e_v) \in (-2.4,2.4) \times (-5,5)$ represents the quantity $\|e_p(t=10)\|$. The peaks (shown in yellow corresponding to points (-2.5,-5) and (2.4,5)) indicate that robotic agents that are moving in opposite directions away from one-another at the initial time are slow to stabilize.}
    \label{fig:value}
\end{figure}

The time instant $t=45$ for injecting the stabilizing $u^*$ is chosen arbitrarily. However through safe-set\footnote{Here, \textit{safe-set} is defined as the set of states from where the stabilizing $u^*$ can drive the states to a given target set in finite time, in presence of adversarially acting delay $d$. } computations using the HJ-reachability approach described earlier, one may inject the stabilizing control in a more principled manner. Figure \ref{fig:value} illustrates this. Part (a) shows the set of initial conditions that enter the region $\|e_p\|\le1.5$ after time $T=10$ seconds when we apply the stabilizing $u^*$ starting at $t=0$. This means that we can choose to `turn off' $u^*$ when the trajectories are in the interior of green \textit{safe-set} and turn it back on for $T$ seconds when the trajectories are near the boundary. The sets in Figure \ref{fig:value} are obtained after a two-step process. First, the optimal stabilizing feedback $u^*(t,x,d)$ is computed by numerically solving the HJI-PDE \eqref{eq:cost_}-\eqref{eq:HJI_}, taking stage cost $l_t=\|e_v\|^2$ and terminal cost $l_T = 0$ (The term $e_v$ is defined as $v_1 - v_2$). Next, this $u^*$ is plugged back into the dynamics to obtain a system that is controlled solely by the delay $d$, and a \textit{HJB}-PDE is solved with stage cost $l_t=0$ and terminal cost $l_T = \|e_p(T)\|$. The resulting optimal value function is visualized in part (b).

\begin{figure}[!htp]
    \centering
    \subfloat[]{  \includegraphics[width=0.8\columnwidth]{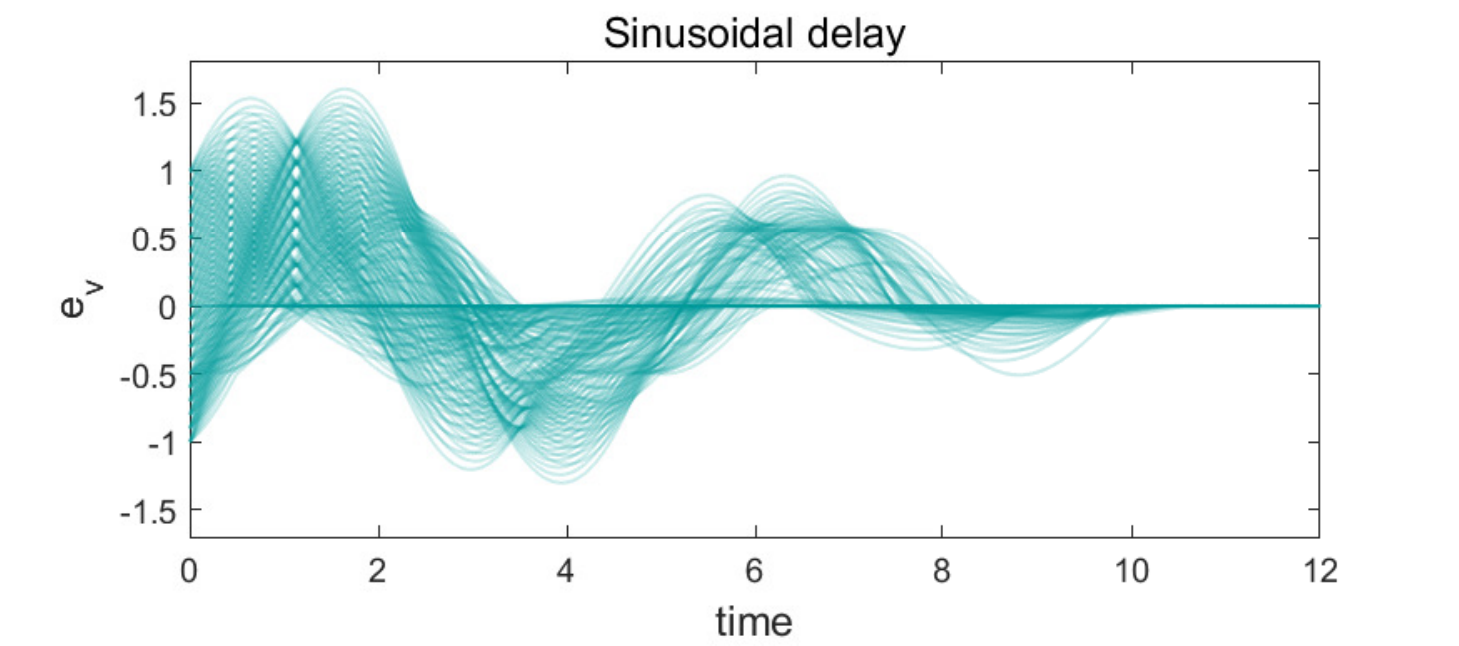}}
    \\
    \vspace{-0.2cm}
    \subfloat[]{
    \includegraphics[width=0.8\columnwidth]{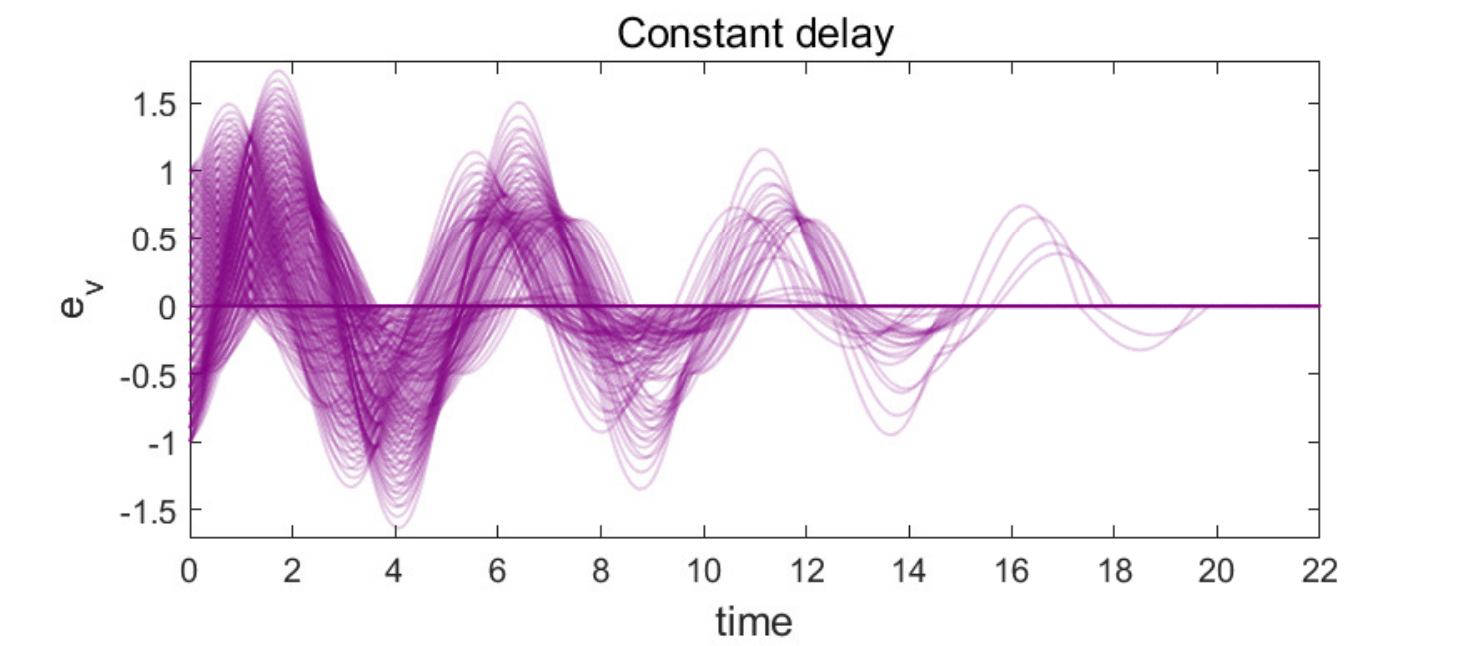}}
    \caption{Under sinusoidal and constant delays, the system trajectories starting from initial points $(e_p(0),e_v(0))$ in the set $ [-1,1] \times [-1,1]$ can be seen to be stabilized. The system parameters are taken to be $k=1, b=0, T^*=0.24$, and the stabilizing optimal control $u^*$ is constrained to be bounded by $u_{max} = 0.4$ and $\|w\| \le 5\|d\|$. One can see that in the constant delay case, the stabilization takes nearly twice as long as the sinusoidal delay case, because the optimal adversarial delays are actually found to be very close to constant for our approximate TDS.}
    \label{fig:trajs}
\end{figure}

Finally, we simulate the closed-loop trajectories of the TDS \eqref{eq:TDS} for different initial conditions and delays as shown in Figure \ref{fig:trajs}. The transients and the steady states show that the cost function that penalizes the quantity $\int_0^T \|e_v(s)\|^2ds$ over a sufficiently long time horizon $T$ indeed stabilizes the system, although the stabilization rate depends on the delay as well as the initial conditions.

    

\section{Conclusion}\label{sec:con}
The existence of time-delays in a system, though largely overlooked in modeling and control, can have adverse effects on stability and performance, even when present in small magnitudes. In this paper, we examine delays as exogenous input signals that adversarially steer the time-delay system into unstable regimes, and consider designing optimal stabilizing controllers for the same. Since these time-varying communication delays enter the dynamics implicitly through the states, analyzing them as a control input is a difficult open problem, which we circumvent through more tractable approximations of the dynamics. We then frame our problem in a differential game setting, where the players are the adversarial time-delay, and the stabilizing control. These players have opposing goals with regards to system stability, over a finite-time horizon. We extract our controller for this zero-sum game by numerically solving the HJI-PDE. We also explore the related notion of HJ-reachability, in context of time-delay systems, and demonstrate computation of a `safe-set' from where our controller can stabilize the system with any bounded delay, in finite time. We present both analysis and numerical experiments to show that our approximation serves well, towards designing stabilizing controllers for the original time-delay system.





\bibliographystyle{ieeetr}
\bibliography{References}

\appendix
\subsection{Proof of Lemma \ref{lemma:approx_error1}}\label{subsec:Proof_approx_error1}
Let us first consider the variables $y_1(t) = \dot{x}_1$ and $y_2 = \dot{x}_2$. Then, $x_1(t-d_1(t)) = x_1(t) - y_1(t)d_1(t) + o(\|d_1\|)$ and $x_2(t-d_2(t)) = x_2(t) - y_2(t)d_2(t) + o(\|d_2\|)$. Thus,
\begin{align*}
& f_1(x_1,x_2(t-d_2),u_1) - f_1(x_1,x_2(t) - y_2(t)d_2(t),u_1) \\
&= f_1(x_1,x_2(t)- y_2(t)d_2(t) + o(\|d_2\|),u_1) \\
&- f_1(x_1,x_2(t) - y_2(t)d_2(t),u_1) \\
&\doteq w_1.
\end{align*}
 Similarly, we can define $w_2$. Then we have,
\begin{eqnarray*}
    \|w_1\| &\le& L_1|o(\|d_2\|)|\\
    \|w_2\| &\le& L_2|o(\|d_1\|)|
\end{eqnarray*}
where $L_1,L_2$ are the respective Lipschitz constants of $f_1,f_2$ with respect to the second argument, that is, $\forall x,a,b,u$ we have $\|f_i(x,a,u)-f_i(x,b,u)\| \le L_i\|a - b\|$, for each $i=1,2$. The constants $L_1,L_2$ are uniform constants due to the derivative boundedness assumption. \qed

\subsection{Proof of Theorem \ref{th:approx_error2}}\label{subsec:proof_approx_error2}

(i) Let us fix initial conditions $x$ and $t$, as well as inputs $u(\cdot)$ and $d(\cdot)$. Let us then define $\tilde{w}(\cdot)$ as 
\[
   \tilde{w}(\tau) \doteq  \dot{\phi}_a(\tau) - f(\phi_a(\tau),\phi_a(\tau) -  \dot{\phi}_ad(\tau),u(\tau)). 
\]
Then, for $w(\tau) = \tilde{w}(\tau)$ and $y(\tau) = \dot{\phi_a}$, $x(\tau) = \phi_a$ trivially satisfies equations \eqref{eq:simple}-\eqref{eq:constraints} and therefore is a solution. In other words,
\begin{equation*}
\phi_a(x,\tau,u(\cdot),d(\cdot))) 
 = \phi_b(x,\tau,u(\cdot),d(\cdot), \tilde{w}(\cdot)),\; t \le \tau \le T , 
\end{equation*}
and therefore the set of trajectories $\phi_a$ is a strict subset of the set of trajectories $\phi_b.$ This concludes the first part of the proof.

(ii) Denote $d_0\doteq w_0 \doteq 0$ for all time.
\small
\begin{align}
    \inf_u J(t,\phi_b(u, d_0, w_0)) &= \sup_{\gamma}\sup_{\delta\in\Delta(t,0)}\inf_u J(t,\phi_b(u,\delta[u],\gamma[u,\delta[u]])) \notag\\
    =&\inf_{\gamma}\sup_{\delta\in\Delta(t,0)}\inf_u J(t,\phi_b(u,\delta[u],\gamma[u,\delta[u]])) \notag\\
    \leq & \inf_{\gamma}\sup_{\delta\in\Delta(t,d_\text{max})}\inf_u J(t,\phi_b(u,\delta[u],\gamma[u,\delta[u]])) \notag\\
    \leq & \sup_{\gamma}\sup_{\delta\in\Delta(t,d_\text{max})}\inf_u J(t,\phi_b(u,\delta[u],\gamma[u,\delta[u]])) \notag\\
    \leq & \sup_{\gamma}\sup_{\delta\in\Delta(t,d_\text{max})} J(t,\phi_b(\tilde{u},\delta[\tilde{u}],\gamma[\tilde{u},\delta[\tilde{u}]])),
    \label{eq:proof_eq_set1}
\end{align}
\normalsize
where $\tilde{u}$ satisfies
$J(t,\phi_b(\tilde{u},d_0,w_0))\leq$$\inf_u J(t,\phi_b(u, d_0, w_0)) +\epsilon$ for $\epsilon>0$.



(iii) We use $\phi_1$ and $\phi_2$ to denote $\phi_b(\tilde{u},\delta[\tilde{u}],\gamma[\tilde{u},\delta[\tilde{u}]])$ and $\phi_b(\tilde{u},d_0,w_0)$ for any $\delta\in\Delta(d_\text{max}), \gamma\in\Gamma$. 
Then, for any $s\in [t,T]$,
\begin{align*}
    \lVert  \phi_1(s) - \phi_2(s)  \rVert \leq \int_t^s  \lVert & f(\phi_1(\tau),\tilde{u}(\tau),\delta[\tilde{u}](\tau),\gamma[\tilde{u},\delta[\tilde{u}](\tau)]) \notag\\
    -& f(\phi_2(\tau),\tilde{u}(\tau),d_0,w_0) \rVert d\tau.
\end{align*}
By Gronwell's inequality and Lipscthiz continuity of $f$, $l_s$, and $l_T$, for all $t\in[0,T]$,
\small
\begin{align*}
    &\lvert J(t,\phi_1)  - J(t,\phi_2) \rvert \notag\\
    \leq &  \left(T \text{exp}(L_f T)-\frac{1}{L_f} (\text{exp}(L_f T)-1) \right) d_\text{max} =C d_{\max}.
\end{align*}
\normalsize
Since the above inequality holds for all $\delta,\gamma$, \eqref{eq:proof_eq_set1} concludes
\small
\begin{align*}
    & \lvert \sup_{\gamma}\sup_{\delta\in\Delta(d_\text{max})} J(\tilde{u},\delta[\tilde{u}],\gamma[\tilde{u},\delta[\tilde{u}]]) - \inf_u J(u, d_0, w_0)\rvert \leq C d_\text{max}+\epsilon.
\end{align*}
\normalsize
and Theorem \ref{th:approx_error2}.
\qed

\end{document}